\documentclass[iop]{emulateapj}

\newcommand{\figscaleone}{\epsscale{1.10}} 

\slugcomment{Submitted to ApJ 10th November 2015; Revised 14th December 2015}

\shorttitle{Cosmic Ray Heating}
\shortauthors{Walker}

\begin{document}

\title{Heating of the Warm ionized Medium by Low-Energy Cosmic Rays}

\author{Mark A. Walker}
\affil{Manly Astrophysics, 3/22 Cliff Street, Manly 2095, Australia}
\email{Mark.Walker@manlyastrophysics.org}

\begin{abstract}
In light of evidence for a high ionization rate due to Low-Energy Cosmic Rays (LECR), in diffuse molecular gas in the solar neighbourhood, we evaluate their heat input to the Warm Ionized Medium (WIM). LECR are much more effective at heating plasma than they are at heating neutrals. We show that the upper end of the measured ionization rates corresponds to a local LECR heating rate sufficient to maintain the WIM against radiative cooling, independent of the nature of the ionizing particles or the detailed shape of their spectrum. Elsewhere in the Galaxy the LECR heating rates may be higher than measured locally. In particular, higher fluxes of LECR have been suggested for the inner Galactic disk, based on the observed hard X-ray emission, with correspondingly larger heating rates implied for the WIM. We conclude that LECR play an important, perhaps dominant role in the  thermal balance of the WIM.
\end{abstract}


\section{Introduction}
At energies $\ga1\,$GeV the cosmic-ray flux in the neighbourhood of the Sun is tightly constrained by direct measurement (e.g. Webber 1998; Olive et al 2014). At lower energies the attenuating effects of the solar wind become increasingly important and much below $100\,$MeV there is little constraint from spacecraft data. Nor is our understanding of cosmic ray acceleration and propagation good enough that we can confidently predict the low-energy fluxes based on theoretical models of those processes. Indirect observational constraints are therefore pivotal. 

As cosmic rays propagate through the ISM, they lose energy through various processes --- ionization, ``pionization'' (for hadrons), and brem\ss trahlung (for leptons) being some of the most important (e.g Olive et al 2014; Padovani, Galli and Glassgold 2009). At low energies ionization dominates in neutral media, and the ionization state of the ISM provides information on the Low-Energy Cosmic Ray (LECR) population (e.g. Hartquist, Black and Dalgarno 1978; van Dishoek and Black 1986; Federman, Weber and Lambert 1996).  Dalgarno (2006) has given an overview of the relevant chemistry and results which have been obtained in dense clouds, diffuse clouds, and the inter-cloud medium. There are difficulties, because ionization balance is influenced by various factors --- e.g. the density, column-density and composition of the cloud, its homogeneity, and the ionizing photon background. Nevertheless in recent years there has been progress from studies of ${\rm H_3^+}$ in diffuse molecular clouds (McCall et al 2003; Dalgarno 2006; Indriolo et al 2007). The particular appeal of ${\rm H_3^+}$ is that the chemistry is simple, so the interpretation of the data ought to be fairly robust (but see Shaw et al 2008). These studies consistently point to large ionization rates -- much higher than can be produced by the known GeV cosmic-rays -- implying high cosmic-ray fluxes at low energies (Indriolo et al 2007; Padovani, Galli and Glassgold 2009; Indriolo, Fields and McCall 2009). In this paper we consider the implications of those results for the LECR heating rate of the diffuse, ionized component of interstellar gas, which we refer to as the Warm ionized Medium (WIM) (Reynolds 2004; Haffner et al 2009).

The LECR interaction which causes ionization of neutrals is just Coulomb scattering -- electrons receive a kick from the electric field of a passing cosmic-ray -- and this process has a larger cross-section when the electrons are free than when they're bound in atoms or molecules. Thus high ionization rates for neutral gas immediately imply large heating rates for ionized gas. We will show that LECR heating alone may suffice to maintain the WIM. That result contrasts with the prevailing view (e.g. Haffner et al 2009; Wood et al 2010; Hill et al 2015) that only UV photons carry enough power to be able to maintain the WIM.

With the data available at present a wide range of spectral forms are possible for the LECR. Consequently it's unclear whether one should think of physically distinct low- and high-energy cosmic-ray components (e.g. with a different acceleration mechanism), or whether they are different aspects of the same phenomenon. Nor is it clear whether the large ionization rates of diffuse molecular gas are caused primarily by cosmic-ray protons or electrons (Padovani, Galli and Glassgold 2009); we therefore consider both possibilities.

Observations of hard X-rays from the inner Galactic disk have also been interpreted in terms of powerful LECR fluxes (e.g. Skibo, Ramaty and Purcell 1996; Valinia et al 2000). Although the inferred LECR spectra, and thus the WIM heating rates, are model-dependent, the X-ray evidence is independent of the molecular ion studies and thus provides a valuable check. Furthemore the X-ray data sample a large volume of the Galactic disk and therefore provide a more representative view of the LECR than local studies afford. 

This paper is structured as follows. In the next section we recap the theory of energy losses via Coulomb collisions of individual LECR, and we apply the theory to both neutral and ionized hydrogen. In section 3 we use the measured ionization rates of diffuse molecular gas to estimate the total Coulomb heating of ionized gas in the solar neighbourhood, largely independent of the uncertainties in the LECR spectrum. Our estimates are comparable to the heating rate needed to maintain the WIM at a temperature $\sim10^4\;$K.  In section 4 we reinforce that point by reference to published fluxes of LECR in the inner Galactic disk, determined from hard X-ray observations. Those studies imply that LECR heating of the WIM could be much larger than our local estimate, and together these results lead us to conclude that LECR may well be the dominant heat source for the WIM.

\section{Energy Losses via Coulomb Collisions}
The interactions between cosmic-rays and molecular hydrogen have been quantified over a broad range of particle energies ($0.1\,$eV to $100\,$GeV) by Padovani, Galli and Glassgold (2009). Olive et al (2014) give a comprehensive review of high-energy processes in different materials. For LECR the dominant energy loss process is Coulomb collisions, treatments of which can be found in many textbooks, e.g. Jackson (1962), Ginzburg and Syrovatskii (1964), and Longair (1981). Our discussion makes use of a simplified calculation in which the expected rate of energy loss of a cosmic ray, of speed $\beta c$, with increasing electron-column, $N_e$, is:
\begin{equation}
-\left\langle{{{\rm d}E}\over{{\rm d}N_e}}\right\rangle_{\!\!coll} = {3\over2}\sigma_T \, {{m_ec^2}\over{\beta^2}} \lambda,
\end{equation}
where $\sigma_T$ is the Thomson cross-section, and
\begin{equation}
\lambda=\ln\left( {{W_{max}}\over{I}}\right)
\end{equation}
is the ``Coulomb logarithm''. Here $W_{max}$ is the maximum amount of energy which can be lost by the cosmic-ray in a single encounter with an electron, and $I$ is the mean excitation energy of the medium associated with individual encounters. In deriving equation (1) the speeds of the target electrons are assumed negligible compared to that of the cosmic-ray. We now proceed to evaluate $\lambda$.

The maximum energy transfer is determined by the kinematics of a head-on collision. For low energy protons (energies $\ll1\,$TeV) the kinematic limit is
\begin{equation}
W_{max}\simeq2 \,\beta^2\gamma^2\,m_ec^2\qquad {\rm(protons)},
\end{equation}
where $\gamma$ is the cosmic-ray Lorentz factor.  For cosmic-ray electrons, on the other hand, one takes
\begin{equation}
W_{max}=\left({{\gamma+1}\over{2}}\right)^{\!\!1/2}\!(\gamma-1)\,m_ec^2 \qquad {\rm(electrons)},
\end{equation}
(Jackson 1962) as the maximum energy transfer.  For both types of primary, the maximum energy transfer is the same for both neutral and ionized gas.

For neutrals, the mean excitation energy is comparable to the ionization energy of the medium. It is difficult to calculate, and in practice it is usually determined by fitting to data. We will consider only the case of a pure  hydrogen target, for which the U.S. National Institute of Standards (NIST) recommends
\begin{equation}
I = 19.2\;\;{\rm eV}\qquad {\rm(neutral\;\,hydrogen)},
\end{equation}
as an appropriate value for both electrons and protons.\footnote{www.nist.gov/pml/data/star/}

If, on the other hand, all the electrons in the target medium are already free then the hydrogen ionization energy is no longer relevant. In that case the mean excitation energy is the quantum of energy associated with Langmuir oscillations in the plasma (Larkin 1959; Ginzburg and Syrovatskii 1964):
\begin{equation}
I = \hbar\omega_p\qquad {\rm(ionized\;\,hydrogen)},
\end{equation}
with $\omega_p$ being the plasma frequency. Numerically we have $\omega_p\simeq 5.6\times10^4\,\sqrt{n_e}\;{\rm rad\,s^{-1}}$, with $n_e$ in ${\rm cm^{-3}}$. Thus if we adopt $n_e\sim0.1\,{\rm cm^{-3}}$, for the WIM (e.g. Haffner et al 2009), then $I\sim10^{-11}\,$eV.  The difference in $I$ values for neutral versus ionized hydrogen is so great that the energy loss rate (1) can be substantially larger for a plasma than for a neutral gas, despite the fact that $I$ only appears in the logarithm. 

%%%%%%%%%%%%%%%%%%%%%%%%%%%%%%%%%%%%%%%%%%
\begin{figure}
\figscaleone
\centerline{\plotone{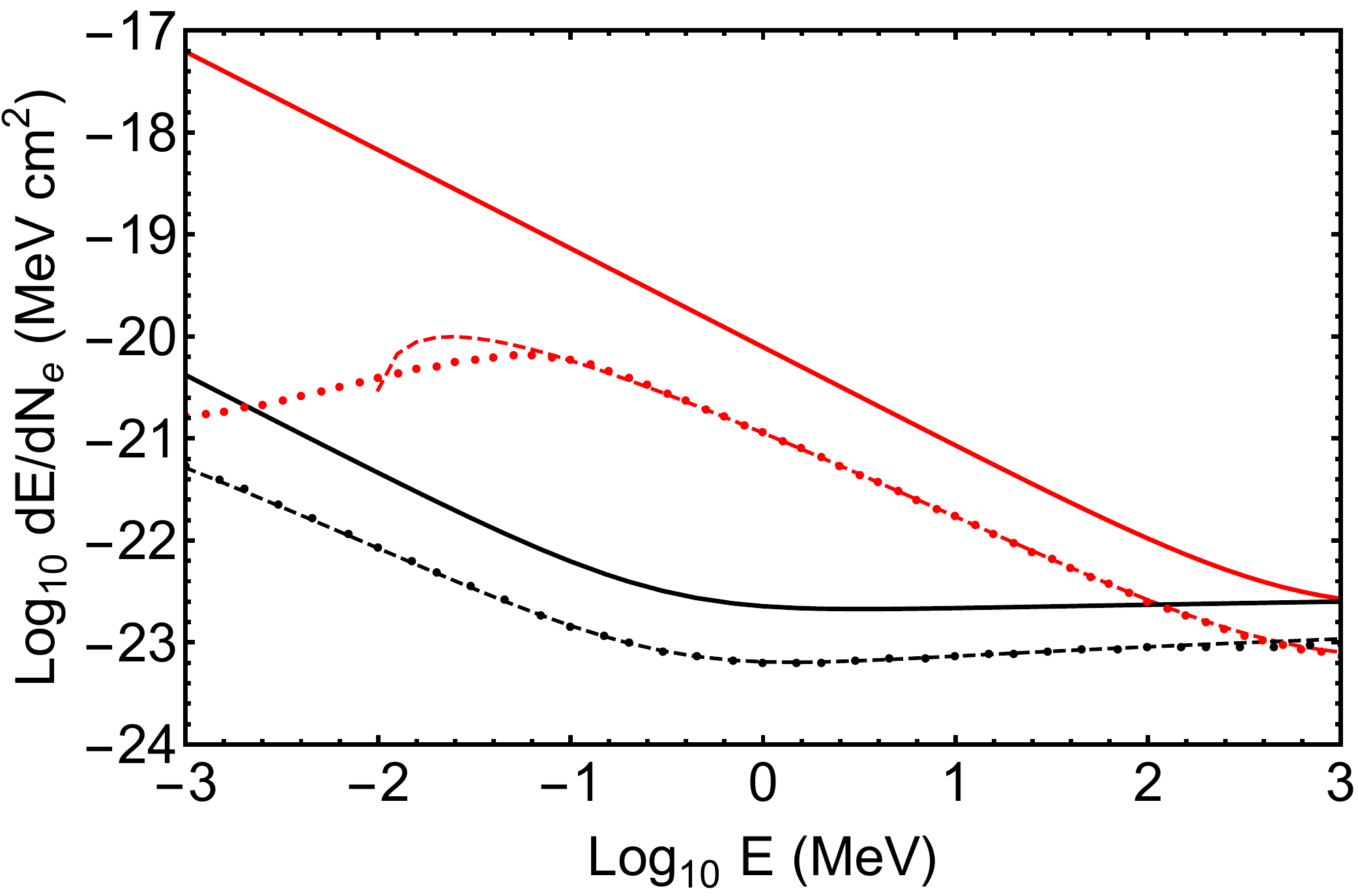}}
\caption{Expected Coulomb losses for cosmic-ray protons (red) and electrons (black) as a function of energy. Points show NIST values for neutral hydrogen in the eSTAR and pSTAR databases. Dashed (solid) lines show the results of our model for neutral (ionized, $n_e=0.1\,{\rm cm^{-3}}$) hydrogen. Above $E\sim300\,$MeV other mechanisms (brem\ss trahlung for electrons, and pion production for protons) become important.}
\end{figure}
%%%%%%%%%%%%%%%%%%%%%%%%%%%%%%%%%%%%%%%%%%

Figure 1 shows the foregoing results in graphical form for cosmic-ray electrons/protons passing through a gas of neutral or ionized hydrogen. Also plotted there are values from the NIST databases$^1$ of collisional stopping power for fast particles in neutral hydrogen --- pSTAR and eSTAR, for protons and electrons, respectively. For LECR protons the differences between pSTAR values and model (1) are better than 10\% for $E\ga100\,$keV. At lower energies we enter the regime where the proton speed is $\beta\la\alpha\simeq1/137$, the speed of electrons bound in hydrogen, so the impulse approximation used to derive equation (1) is no longer valid. For cosmic-ray electrons this circumstance is not reached until $E\la100\,$eV, and there is agreement between the predictions of model (1) and the eSTAR values to better than 5\%\ from $1\,$keV to $100\,$MeV. Because our model works well for neutral gas, we expect it to provide a good approximation to the Coulomb losses in ionized gas providing only that the thermal speeds of the target electrons are negligible. For the WIM we are interested in plasma temperatures of order $10^4\,{\rm K}$, which limits the applicability of model (1) to protons of energy $\ga1\,{\rm keV}$ and electrons of energy $\ga1\,{\rm eV}$, and thus includes the full range of energies shown in figure 1.

In subsequent calculations we use the NIST values of Coulomb losses for neutral hydrogen, and for ionized hydrogen we use model (1) with an assumed density of $n_e=0.1\,{\rm cm^{-3}}$. We neglect all other loss processes, so our model only applies to LECR ($E\la100\,$MeV). We now turn to the calculation of the total Coulomb loss rate for LECR in the solar neighbourhood.

\section{Local Heating by LECR}
It has already been established (Padovani, Galli and Glassgold 2009; Indriolo, Fields and McCall 2009) that flat LECR spectra cannot explain the observed ionization rates of diffuse molecular gas: the spectrum must rise substantially below $100\,{\rm MeV}$. In this paper, therefore, we consider only steep LECR spectra, for which the total Coulomb losses are dominated by the low energy particles in the spectrum. Unfortunately the detailed spectral shape of LECR in the solar neighbourhood is not known. Nor, even, do we know whether protons or electrons are principally responsible for the ionization of diffuse molecular gas (Padovani, Galli and Glassgold 2009). Furthermore the low-energy spectrum of either species is modified during propagation. We can nevertheless estimate LECR heating in ionized gas simply by scaling from the measured ionization rate in neutral gas, as follows. 

%%%%%%%%%%%%%%%%%%%%%%%%%%%%%%%%%%%%%%%%%%
\begin{figure}
\figscaleone
\centerline{\plotone{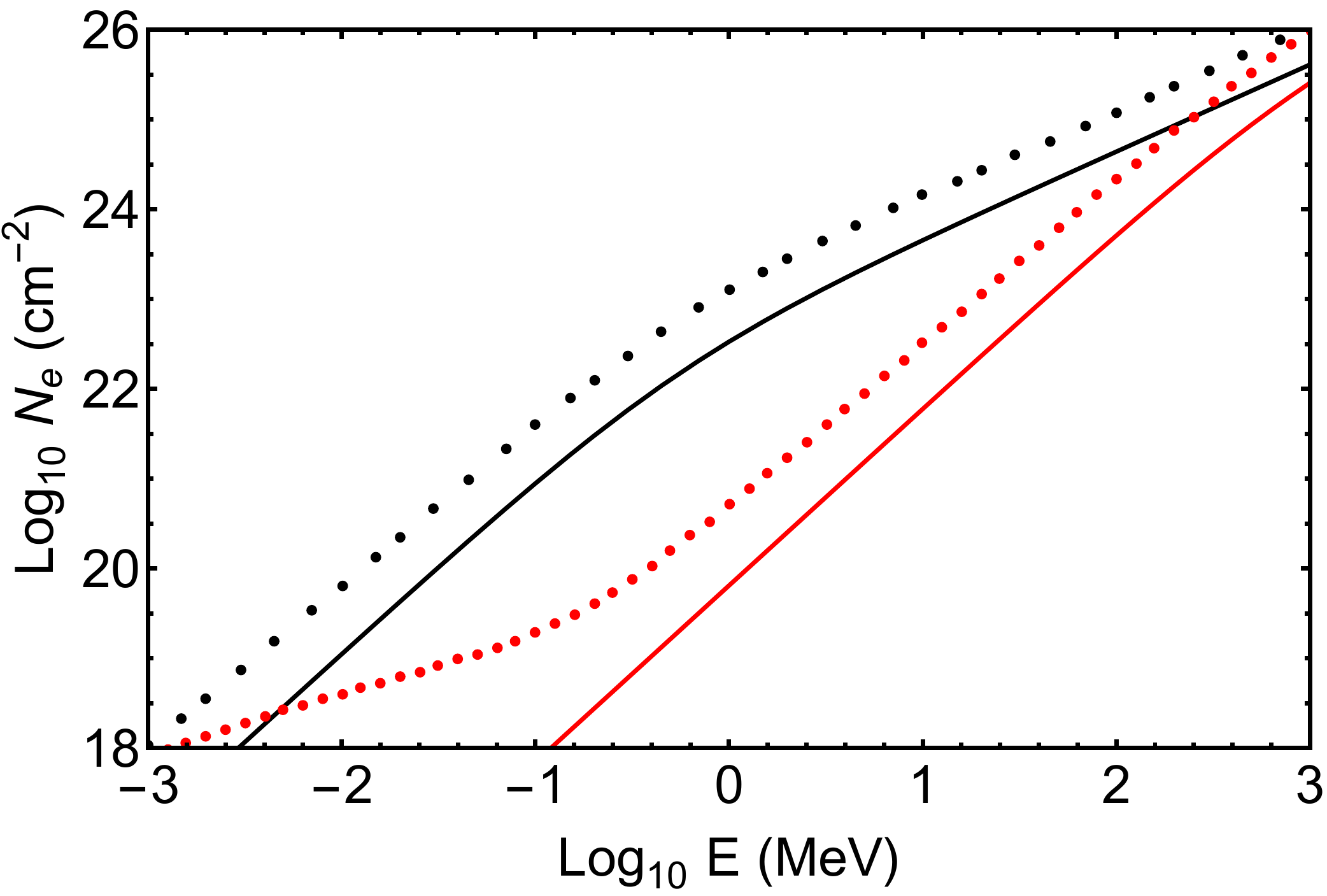}}
\caption{The particle range in ionized (lines) and neutral hydrogen (dots) for energetic electrons (black), and protons (red), computed in the ``continuous slowing down approximation'', assuming only Coulomb losses --- as per the NIST databases eSTAR and pSTAR, respectively. Neglect of brem\ss trahlung (electrons), and pion production (protons) means that ranges are significantly overestimated at $E\ga300\,$MeV. Note that cosmic-rays propagate diffusively, in a random walk, and ``range'' is the column measured along the actual path taken.}
\end{figure}
%%%%%%%%%%%%%%%%%%%%%%%%%%%%%%%%%%%%%%%%%%

Equation (1) describes the rate at which  a single cosmic-ray loses energy via Coulomb collisions. For a particle spectrum $j(E)$ (${\rm cm^{-2}\,s^{-1}\,sr^{-1}\,MeV^{-1}}$), the total rate at which energy is lost, i.e. the total dissipation rate, $H$, is just the integral of the stopping power over the spectrum and solid-angle, $\Omega$:
\begin{equation}
H=-\int\!\!\int\!{\rm d}\Omega\,{\rm d}E \left\langle{{{\rm d}E}\over{{\rm d}N_e}}\right\rangle_{\!\!coll} j(E).
\end{equation}
Now figure 1 shows that, except for the lowest energy protons ($E<0.1\,$MeV), the ratio of stopping powers in neutral and ionized media varies only slowly with particle energy. So if we have a rough estimate of the  particle energies which dominate the total dissipation, we can evaluate the ratio of stopping powers in the two media at that energy, and that should give us a good estimate of the ratio of the total dissipation rates.

%%%%%%%%%%%%%%%%%%%%%%%%%%%%%%%%%%%%%%%%%%
\begin{figure}
\figscaleone
\centerline{\plotone{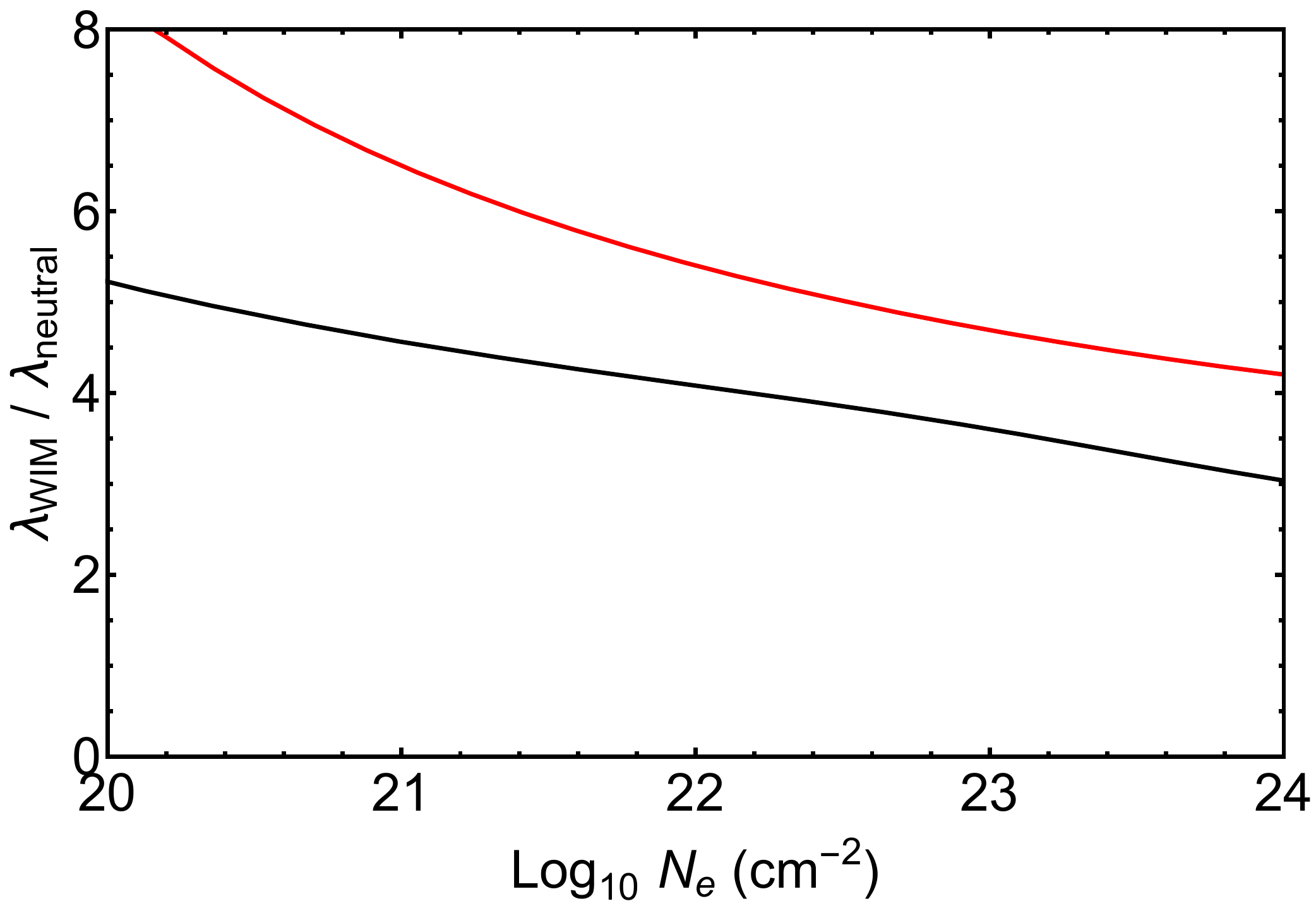}}
\caption{The ratio of stopping power in the WIM to that in neutral hydrogen, for protons (red) and electrons (black), as a function of the expected particle range in neutral gas. For the clouds used by Indriolo et al (2007) to determine LECR ionization rates, the typical column encountered by LECR is $N_e\sim10^{22}\,{\rm cm^{-2}}$.}
\end{figure}
%%%%%%%%%%%%%%%%%%%%%%%%%%%%%%%%%%%%%%%%%%

In neutral gas the dissipation rate can be gauged by the primary ionization rate, $\zeta$. Indriolo et al (2007) summarised data on the ionization rate of diffuse molecular gas in the solar neighbourhood, arriving at the conclusion that $\log_{10}\zeta\,({\rm s^{-1}\;proton^{-1}})\simeq-15.7\pm0.5$. And for fast electrons in molecular hydrogen Dalgarno, Yan and Liu (1999) determined the mean energy required per ion pair to be $\langle E_{ion}\rangle\simeq40\,$eV. We adopt this value for cosmic-ray protons also. Thus in diffuse molecular gas in the solar neighbourhood the LECR dissipation rate is 
\begin{equation}
\log_{10}H=\log_{10}\zeta+\log_{10}\langle E_{ion}\rangle=-25.9\pm0.5,
\end{equation}
with $H$ in units of ${\rm erg\,s^{-1}\,proton^{-1}}$.

The clouds considered by Indriolo et al (2007) have a broad range of column-densities and sizes; we adopt values of $N_e\sim3\times10^{21}\,{\rm cm^{-2}}$, and $L\sim4\,{\rm pc}$, respectively, as representative. As a result of scattering by magnetic irregularities, cosmic-ray transport is diffusive and a cosmic-ray which traverses an entire cloud may encounter a total column-density much larger than the line-of-sight value given by Indriolo et al (2007). There is currently insufficient data on LECR to permit useful constraints on the quantitative characteristics of their diffusion. For GeV cosmic-rays, values of the diffusion coefficient of $D\sim4\times 10^{28}\,{\rm cm^2\,s^{-1}}$ typically provide a good match to the data (Strong, Moskalenko and Ptuskin 2007), corresponding to a mean-free-path to large-angle scattering of $3D/c\sim1\,{\rm pc}$. We assume that this mean-free-path also holds for LECR, implying that the column-density encountered by a cosmic-ray traversing a typical cloud in the sample of Indriolo et al (2007) is $N_e\sim10^{22}\,{\rm cm^{-2}}$. Because the LECR spectrum is steep, the total dissipation rate in a typical cloud should therefore be dominated by particles whose range is  $\sim10^{22}\,{\rm cm^{-2}}$.

Figure 2 shows the ranges of electrons and protons in neutral hydrogen, determined by integrating the reciprocal of the stopping powers shown in figure 1. From figure 2 we can see that particle ranges $\sim10^{22}\,{\rm cm^{-2}}$ correspond to protons of energy $\sim5\,$MeV, and electrons of energy $\sim0.2\,$MeV, and it is particles of about these energies which we expect to make the main contribution to the ionization rates given by Indriolo et al (2007). As can be seen in figure 1, such particles have stopping powers which are, respectively, $5.4$ and $4.1$ times greater in the WIM than in neutral gas --- see also figure 3, where the ratios of stopping power are plotted as a function of particle range in neutral gas.\footnote{Using more sophisticated models for the stopping of fast particles, from Olive et al (2014), alters the stopping power ratios by less than 1\%\ over the full domain of figure 3.} In the case of ionized gas, all of the power dissipated from LECR goes into heat, so the estimated heating rate of the WIM in the solar neighbourhood (in ${\rm erg\,s^{-1}\,proton^{-1}}$) is:
\begin{equation}
\log_{10}H_p\simeq -25.2\pm0.5 
\end{equation}
if the LECR are predominantly protons, and 
\begin{equation}
\log_{10}H_e\simeq -25.3\pm0.5 
\end{equation}
if the LECR are predominantly electrons.

The midpoints of estimates (9) and (10)  are both below the heating rate needed to maintain the WIM at $10^4\,$K, which is $\log_{10}H_{WIM}\simeq -25.1$ at density $n_e=0.1\,{\rm cm^{-3}}$ (Reynolds 1990). However, neither estimate is far below the requisite power. Even at the lower end of the indicative range LECR are expected to contribute 20-25\% of the WIM's power, and at the upper end of the range the WIM can be maintained against radiative cooling by LECR heating alone, regardless of the nature of the primaries.

The estimates (9) and (10) are valid for the WIM in the neighbourhood of the Sun, with ``neighbourhood'' defined by the sample of diffuse molecular clouds in Indriolo et al (2007). The distances of the target stars in that sample range up to a few ${\rm kpc}$, but most are below $1\,{\rm kpc}$ and so we expect the intervening diffuse clouds to lie at distances of order a few hundred parsec from the Sun. As the target stars lie at low Galactic latitudes, this distance estimate is in effect a radius within the Galactic plane. The estimates (9) and (10) should, however, be valid at least as far off the plane as LECR can diffuse. If, as assumed above, the mean-free-path to scattering of LECR is $\sim1\,{\rm pc}$, then electrons of energy $\sim0.2\,$MeV and protons of energy $\sim5\,$MeV should be able to diffuse a distance $\sim10^2\,{\rm pc}$ through the WIM. Absent a reliable model of acceleration and propagation for LECR, we cannot say whether or not the estimates (9) and (10) are appropriate at larger distances from the plane.

\subsection{The influence of helium}
The foregoing calculations neglect the presence of He in diffuse molecular clouds and the WIM, whereas we expect helium to be roughly 25\%\ by mass in both environments.  We now consider how our estimates (9) and (10) would change if we accounted for the presence of helium. 

First we note that the input value of ionization rate of diffuse H$_2$ and the associated LECR dissipation rate are both unchanged by the presence of helium --- precisely because they are specific to hydrogen. Similarly, the WIM power estimates are normalised to the free-electron content, and in the WIM it appears that helium is predominantly neutral  and hydrogen is predominantly ionized (Reynolds 2004), thus both the emission line data and our own estimated power are effectively normalised to the number of hydrogens.

There are, however, two small effects which we must account for. First, there will be an additional contribution to the heating of the WIM arising from LECR ionization of helium atoms. Helium contributes one additional electron for every six from hydrogen, but that electron is bound, and we calculated that the associated LECR dissipation rate is smaller by factors of 5.4 and 4.1 for LECR protons and electrons, respectively, in the case of electrons bound in hydrogen. Because of the higher ionization energy of helium, its stopping power per electron is roughly 10\%\ smaller than hydrogen (see the NIST databases), so the contribution of helium to the LECR dissipation rate in the WIM is only one part in 36 for LECR protons, and one part in 27 for LECR electrons. Furthermore, in the case of cosmic-ray ionization of neutrals, only a small fraction of the dissipated energy goes into heating the gas -- Dalgarno, Yan and Liu (1999) estimated that fraction to be 0.16 in the case of helium -- with the result that the presence of helium is expected to increase the LECR heating rate in the WIM by $0.4$\%\ (protons) to $0.6$\%\ (electrons). 

The second effect that we need to quantify is the increase in column-density of the diffuse molecular clouds used to measure the LECR ionization rate. The total electron column increases by approximately 16\%\ when we allow for the presence of helium, so the ratio of dissipation rates in ionized and neutral media must  decrease accordingly. However, that ratio is only weakly dependent on the column-density (figure 3), and the implied change in WIM heating is thus a decrease of only about 1\%. 

The net result of these two effects is a decrease in the expected WIM heating by an amount of order 0.5\%. This change is small in comparison with the uncertainty in the estimates (9) and (10).

\section{LECR Heating in the Inner Galaxy}
Strong LECR fluxes have been inferred in a number of studies of the hard X-ray emission of the inner Galactic disk. Although X-ray emission can arise from either electrons (brem\ss trahlung) or protons (inverse-brem\ss trahlung), models in which a large fraction of the hard X-ray emission is attributed to LECR ions are disfavoured because they tend to predict too much nuclear gamma-ray line emission from the ISM, and too much beryllium production via spallation (Tatischeff, Ramaty and Valinia 1999; Valinia et al 2000). At present there is no consensus on the shape of the LECR electron spectrum; nevertheless, high ionization and dissipation rates are a common feature of the models (Skibo and Ramaty 1993; Skibo, Ramaty and Purcell 1996; Valinia and Marshall 1998; Valinia et al 2000).

In some cases authors quote the ionization rate implied by their models. For example: Skibo and Ramaty (1993) modelled the  $0.03-1000\,$MeV hard X-ray/gamma-ray data from the central radian of the Galaxy, and estimated a primary ionization rate of $\zeta\sim10^{-15}\,{\rm s^{-1}}$ due to the LECR electrons in their model.  This estimate is an average over the inner Galactic disk, so it applies to the ISM throughout that region. It is five times larger than the local value of $\zeta$ from Indriolo et al (2007),  used in \S3. If the shape of the LECR spectrum is similar to that local to the Sun, the implied heating rate for the WIM is then five times larger than the estimate given in equation (10).

The high ionization rate deduced by Skibo and Ramaty (1993) is associated with the low-energy end of their model electron spectrum, where fluxes must be large if the $30\,$keV X-rays are primarily from brem\ss trahlung. Skibo, Ramaty and Purcell (1996) argued in favour of a large brem\ss trahlung contribution to the diffuse Galactic emission down to $\sim10\,$keV, which pushes the LECR ionization rate even higher. They inferred a primary ionization rate of $1.6\times10^{-14}\,{\rm s^{-1}}$, suggesting a heating rate that is 80 times larger than the estimate in equation (10).

Decreasing the contribution made by LECR electrons to the observed $10\,$keV emission naturally leads to lower estimates of the dissipation rate. Valinia et al (2000) modelled the $<10\,$keV X-ray spectrum of the inner Galaxy with a combination of thermal plasma and non-thermal electrons. They did not quote an ionization rate, but they did give their model LECR electron spectrum:
\begin{equation}
J_e=1.7\times10^{-6}\left({E\over{E_0}}\right)^{\!0.3}\exp(-E/E_0), 
\end{equation}
in ${\rm cm^{-3}\,sr^{-1}\,MeV^{-1}}$, with $E_0=0.09\,$MeV. The corresponding heating rate $({\rm erg\;s^{-1}\,proton^{-1}})$ for the WIM can be evaluated from equation (7), wth $j_e=J_e\,\beta c$, it is:
\begin{equation}
\log_{10}H_e\simeq -23.5,
\end{equation}
i.e. roughly 60 times larger than the local estimate (10).

The approach of Valinia et al (2000) simultaneously modelled the properties of the Fe K emission line, which exhibits a fluorescent component at $6.4\,$keV, and the X-ray continuum shape. A recent study of hard X-ray line and continuum emission from the inner few degrees of the Galactic disk concluded that LECR protons explain the data better than LECR electrons (Nobukawa et al 2015). Their preferred model proton spectrum is steep and has a large energy density, and the implied heating rate of the WIM is three orders of magnitude larger than our local estimate (9). That is so large that it may be difficult to identify a power source able to maintain such a spectrum in steady state. For the present discussion, though, the important point is that an explanation of the hard X-rays which relies on LECR protons, rather than electrons, does not evade the implication of a large heating rate for the WIM.

We can summarise this section in the following way: if a significant fraction of the observed hard X-ray emission from the inner Galaxy is due to LECR, then those LECR heat the WIM at a much greater rate than the local estimates given in \S3.

\section{Discussion}
Various processes have been contemplated as heat sources for the WIM, including: photoionization (e.g. Mathis 1986); photoelectric heating by dust grains (Reynolds and Cox 1992; Weingartner and Draine 2001); dissipation of hydromagnetic wave energy (Minter and Spangler 1997; Wiener, Zweibel and Oh 2013)\footnote{Wiener, Zweibel and Oh (2013) considered waves generated as a result of cosmic-ray streaming.}; and, of course, the Coulomb collisions of cosmic-rays. Of these, photoionization and LECR heating are of special interest because they may also be able to account for the ionization of the WIM.  Lyman continuum photons and LECR are similar in that both appear able to supply the necessary power, and ultimately they have a common origin in young, massive stars (which are the progenitors of supernovae). But LECR and UV photons differ profoundly in the way they propagate through the Galaxy. The large absorption cross-section of neutral hydrogen to the Lyman continuum ($\sim10^{-18}\,{\rm cm^2}$) means that these UV photons are not expected to travel far from their source. By contrast cosmic-rays are able to penetrate much larger columns of gas, particularly neutral gas, depending on their energy (see figure 2). Because of this difference, LECR are more readily able to supply the widespread power needed to explain the WIM. 

Many authors have argued that non-uniform gas density in the ISM can lead to Lyman continuum photons propagating to large distances from the source in some directions (e.g. Dove and Shull 1994; Wood et al 2010), making it possible to explain the broad spatial distribution of the WIM in terms of photoionization. However, the spectrum of the WIM is also a problem for photoionization models: the observed forbidden-line ratios are distinctly different from those exhibited by ``classical'' HII regions (which are certainly photoionized), consistent with the WIM being significantly hotter (Haffner, Reynolds and Tufte 1999). Moreover the observed trends in line-ratios with gas density suggest that WIM heating cannot be solely the heating associated with ionization (Haffner, Reynolds and Tufte 1999; Reynolds, Haffner and Tufte 1999). Because LECR heat the free electrons themselves, as well as contributing heat by ionizing neutrals, the observed forbidden-line ratios pose no special problem for models in which LECR dominate the heating of the WIM. 

The potential importance of LECR in determining the state of interstellar gas has long been appreciated. Indeed theoretical models in which LECR are the principal source of heat and ionization actually predate the discovery of the WIM (Spitzer and Tomasko 1968; Field, Goldsmith and Habing 1969; Goldsmith, Habing and Field 1969). The main difficulty with such models has always been the very large uncertainty in LECR flux. Substantial uncertainty remains, and there are indications that LECR fluxes vary substantially from place to place, even in the solar neighbourhood (Indriolo and McCall 2012), but it now seems likely that LECR are playing an important role in the WIM and detailed models specific to that context would be valuable. In addition to the key issue of whether or not one can build an acceptable model of the WIM based on LECR heating and ionization, there is the prospect that the available nebular diagnostics for the WIM could provide some new insights into the LECR population.

Assuming Case B conditions, with a temperature of $10^4\,{\rm K}$, and a density $n_e\sim0.1\,{\rm cm^{-3}}$, the recombination rate in the WIM is $\sim3\times10^{-15}\,{\rm cm^{-3}\,s^{-1}}$. Reynolds et al (1998) showed that [OI]$\;\lambda6300$ emission from the WIM is weak, and inferred that the density of atomic hydrogen in the WIM, $n_a$, must be small compared to that of ionized hydrogen, so $n_a\la0.1\,{\rm cm^{-3}}$. Accounting for secondary ionizations (e.g. Cravens and Dalgarno 1978), the total LECR ionization rate is approximately $1.7\,n_a\zeta$, so for this rate to balance recombinations we must have $\zeta\ga 2\times10^{-14}\,{\rm s^{-1}}$ --- i.e. at least as large as the value inferred by Skibo, Ramaty and Purcell (1996), for LECR electrons in the inner Galaxy (\S4). This result is not unreasonable. As already noted, LECR fluxes may vary significantly from place to place in the Galaxy, and those regions where the ionized fraction is highest are a priori likely to be those where the ionizing fluxes are greatest.

\section{Conclusions}
The measured ionization rates in diffuse molecular clouds, local to the Sun, imply that LECR fluxes are high --- high enough that they may be able to maintain the WIM at $10^4\,$K without the need for any other source of heat. LECR also lead to diffuse, hard X-ray emission from the ISM, and X-ray observations of the inner Galactic disk provide independent evidence of strong LECR fluxes. The WIM heating rates implied by the X-ray data are model-dependent, but they're consistently even higher than our local estimate from the molecular ion data. Although large amounts of power flow in both Lyman continuum photons and LECR, the spatial distribution and spectral properties of the WIM favour LECR as a more natural explanation of the requisite heating. 

\acknowledgments
Thanks to Tony Bell, Don Melrose and Mark Wardle for their help.

\end{document}